\documentstyle[12pt,aasms4]{article}

\def\BmV0{\mbox{(B-V)$^{\rm o}$}}
\def\VmK0{\mbox{(V-K)$^{\rm o}$}} 
\def\MV0{\mbox{M$_{\rm V}^{\rm o}$}}

\def\Msun{\mbox{M$_{\odot}$}}

\def \lta {\mathrel{\vcenter
     {\hbox{$<$}\nointerlineskip\hbox{$\sim$}}}}
\def \gta {\mathrel{\vcenter
     {\hbox{$>$}\nointerlineskip\hbox{$\sim$}}}}

\begin{document}

\lefthead{R-process in low-mass Stars}
\righthead{R-process in low-mass Stars} 

\title {R-Process in Collapsing O/Ne/Mg Cores}

\author{J. Craig Wheeler,\altaffilmark{1}
John J. Cowan,\altaffilmark{2}
and Wolfgang Hillebrandt\altaffilmark{3}}
\begin {center}
wheel@astro.as.utexas.edu,
cowan@mail.nhn.ou.edu,
wfh@mpa-garching.mpg.de
\end {center}

\vskip .5truein

\altaffiltext{1}{Department of Astronomy and McDonald
 Observatory, University of Texas, Austin, TX 78712}
\altaffiltext{2}{Department of Physics and Astronomy,
University of Oklahoma, Norman, OK 73019 }
\altaffiltext{3}{Max-Planck-Institut f\"ur Astrophysik, 
D-85740 Garching, Germany }

\begin{abstract}

Several circumstantial arguments point to the formation of the 
third {\it r}-process peak at A$\sim$190, near platinum, in stars
of mass of $\sim$ 8-10 \Msun : 1) The delayed production of europium
with respect to iron imposes a time scale that restricts the progenitor
stars to M$\lta$10 \Msun ; 2) the {\it r}-process demands a dominant robust
mechanism at least for Z$\ge$56, barium and above, 
since the relative abundance pattern of those {\it r}-process elements 
in the low-metallicity stars, [Fe/H] $<$ -2, is consistent
with the solar pattern;  
3) stars of $\sim$ 8-10 \Msun\ produce nearly identical degenerate
O/Ne/Mg cores that collapse due to electron capture; and 4)
the resulting low-mass cores may produce both an {\it r}-process
in a prompt explosion and a subsequent {\it r}-process in a neutrino
driven wind.  
A special case of the {\it r}-process singles out low entropies for initial
Y$_e\sim$$\bar{Z}$/A, where $\bar{Z}$ is the mean atomic
number of the seed nuclei and A is the atomic weight of
the target. For $\bar{Z}\simeq$35 and Y$_e\simeq$0.18 the A$\simeq$190 
peak results in a natural way.   
The prompt explosion of an O/Ne/Mg core yields 
low entropy, S$\sim$15, and low electron fraction, Y$_e\simeq$0.2,
and hence may produce a reasonable {\it r}-process peak at A$\simeq$190,
as well as all of the {\it r}-process elements with Z$\ge$ 56. 
The possible differences in the $\nu$-driven wind
and associated {\it r}-process due to the low-mass neutron stars expected
in this mass range are also discussed.
\end{abstract}


\section{Introduction}

The {\it r}-process defined by the classic paper of Burbidge, Burbidge, 
Fowler, \& Hoyle (1957) remains
one of the most intriguing in the study of nucleosynthesis.  
The actual {\it r}-process site has still not been identified, 
although supernovae are strongly suspected. 
Cowan, Thielemann, \& Truran (1991), Mathews, Bazan, \& Cowan (1992) and
Meyer (1994) discuss possible sites.
Recent developments in both observations and theory have brought this process
into
special relief.   

Of particular importance have been studies of the abundances of
{\it r}-process and other elements in halo stars of low metallicity
(\cite{spi78,gil88,sne96,bur97} and references therein).
The {\it r}-process elements cannot be synthesized internally in low-mass
stars.  The
presence of these elements in the oldest Galactic stars
(i.e. early in the history of the Galaxy) points to rapidly
evolving and hence massive progenitors for the {\it r}-process.
Two important conclusions have recently come into focus.  

One is the evidence that the production of iron at low metallicity, and 
hence early in the history of the Galaxy, occurred substantially
before the production of the {\it r}-process element europium.  
At the lowest metallicities  the stellar abundance of europium 
begins to deviate from a strict scaling with the abundance of iron
(\cite{gil88}). 
This suggests that the stars that produce iron must have evolved before the
stars
that produced the europium.  
The data were interpreted by Mathews, Bazan, \& Cowan (1992)  
as an evolutionary time-delay for {\it r}-process production
by $\sim 10^8$ yr that is related to the masses of the 
{\it r}-process progenitors. While noting there were uncertainties
in their model calculations, Mathews et al.  
found that a model in which iron came from massive stars,
$\gta$ 15 \Msun, and europium came from stars of $\sim$ 8-10 \Msun\
gave a good fit to the stellar abundance data.

An equally striking conclusion arises from 
recent space-based observations (\cite{cow96,sne97}) in conjunction with
ground-based observations (\cite{cow95,sne96,cow97a})  
yielding abundance data from just above the
second {\it r}-process peak (A $\sim$ 130) to beyond the third {\it r}-process
peak (A $\sim$ 190) 
at low metallicity, [Fe/H] ranging from -2 down to of order -3.
The detailed distributions of the {\it r}-process elements at
these low metallicities, at least for the elements with Z $\ge$ 56, 
is consistent with a scaled solar system distribution.
This strongly implies that there has been a single
dominant {\it r}-process working throughout the history of the Galaxy.
Although there is some evidence for more than one {\it r}-process mechanism,
the conclusion is that the dominant process must be rather robust,
not an integration over broad ranges of initial
seeds and other physical conditions.  

On the theoretical side, 
explorations of {\it r}-process conditions also suggest only a 
limited number of environments could be responsible for producing an 
abundance pattern consistent with the solar one (\cite{thi97}).
The general physical conditions necessary to create the 
{\it r}-process have been continually refined 
(Takahashi et al. 1994; Fuller \& Meyer 1995; Hoffman et al. 1996). 
The interesting physics of $\nu$-driven winds has been proposed
and explored in greatly increasing detail   
(Woosley \& Hoffman 1992; Howard et al. 1993; Takahashi, Witti, \& Janka 1994;
Qian \& Woosley 1996; Hoffman, Woosley \& Qian 1997). 
The $\nu$-driven wind, while based on very enticing physics, has not
yet proven to give adequate results, especially for
the third {\it r}-process peak (Takahashi et al. 1994).  It is not clear 
whether this is an intrinsic problem with the physics of the process
or whether it has been applied in an inappropriate context.
In particular, most of the focus of the work on the {\it r}-process
has been on collapsing iron cores of massive stars.  This is
because these models are directly relevant to SN 1987A
and most of the collapse models currently being studied
are representative of this progenitor mass, $\sim$ 15 -- 20 \Msun .
Given the evidence from the europium and the arguments we
present here, this may prove to be the wrong mass range for
the {\it r}-process.  

Aside from the evidence from europium that stars of mass less
than $\sim$ 10 \Msun\ are the relevant place to look for the {\it r}-process,
these stars are of great intrinsic interest.  
It is {\it a priori} clear that, if they explode, 
stars in the mass range $\sim$ 8-10 \Msun\ 
will be a substantial fraction of all supernovae
and that they will undergo core collapse that is significantly
different than that of the iron cores of more massive stars.
This is because just above $\sim$8 \Msun, the limit to produce
degenerate C/O cores, stars produce degenerate O/Ne/Mg cores (Nomoto 1984).  
The upper limit to this process is somewhat uncertain,
but it may be about 12 \Msun, above which stars produce iron cores 
(Nomoto 1987).

There are two significant properties of the final evolution
of stars in the mass range $\sim$ 8-10 \Msun.  
First, the 
evolution of the cores
should be virtually identical.  This alone makes them an
enticing site for a robust {\it r}-process, independent of the evidence
from europium. At higher masses stars undergo non-monotonic
core evolution (Marom \& Barkat 1990;  Woosley \& Weaver 1995), 
with various core masses and various envelope masses and structures.
This is not conducive to production of a single robust {\it r}-process,
even if a compelling argument for the $\nu$-driven wind could
be made for the massive stars.  The second attribute of these
low-mass stars is that they collapse by electron capture.
This gives their cores an intrinsically different (lower) initial
$Y_e$. They also have a lower core mass, a proper mass of 
$\sim$ 1.4 \Msun\, and so they will form neutron stars with
significantly lower mass but  larger radii.  Finally, 
the cores burn oxygen during the collapse, a process that does not
halt the collapse, but can slow the infall and alter the collapse adiabat 
(Hillebrandt, Nomoto, \& Wolff 1984; but also see Burrows \& Lattimer 1985). 
The locus of oxygen burning can define an even smaller collapsing iron
core, $\sim$ 1.2 \Msun, in the calculation of Hillebrandt et al.
The smaller mass is intrinsically better able to produce a
robust explosion because the envelope is more loosely bound with
no dense mantle and is separated from the core by a steep density gradient.
Once a strong shock is launched, stalling in the mantle is not an
issue. This is the reason why such collapsing cores
remain the only potential candidates for prompt explosions. In any case,  
these stars must produce an interesting class of 
neutron stars if they proceed to collapse and they clearly
have properties of relevance to the general issue of the
{\it r}-process.

\section{Collapse of O/Ne/Mg Cores}

The collapse of O/Ne/Mg cores has received relatively little
attention in the gravitational collapse community.  One
especially relevant work in this regard is the paper by
Hillebrandt, Nomoto, \& Wolff (1984) in which the electron-capture-induced
collapse of such a core was followed in some detail.  This degenerate
core did not reach the Chandrasekhar mass because of the
destabilizing effects of electron capture.  
Rather, it had a proper mass of 1.38 \Msun\ at the time of collapse
(Nomoto 1987).  A few other studies have been done of collapsing 
degenerate cores of 1.4 \Msun.  C/O cores were investigated
by Baron et al. (1987b), Woosley \& Baron (1992), and 
Colgate, Fryer, \& Hand (1997) and O/Ne/Mg
cores by Baron et al. (1987a) and Wilson \& Mayle (1988), 

In the calculation of Hillebrandt, Nomoto, \& Wolff (1984), the burning
of oxygen is found to appreciably alter the velocity and entropy of
the infall.  In these calculations, a prompt explosion occurred, ejecting 
the portion of the core beyond the standing oxygen-burning front,
$\sim$ 0.2 \Msun.  This prompt shock ultimately determined the entropy 
of the ejecta which
was about 15 in units of k, the Boltzmann constant, for the mass
shells of interest here. The electron fraction is
rather low in the center of the core during the collapse,
with a value of Y$_e\sim0.36$ due to the preceding electron captures.
This has some effect on the formation of the neutron star, but
cannot directly affect the nucleosynthesis since this matter will
remain in the core of the neutron star.  Rather, Hillebrandt et al.
note, the inner 0.02 \Msun\ of ejecta that do escape have
neutron to proton ratios of 4 to 5 and hence $Y_e\sim 0.16 - 0.2$.

These very low values of $Y_e$ are extreme by most current
considerations of core collapse and {\it r}-process models, but they
do arise rather naturally in the specific context of prompt
explosions in collapsing O/Ne/Mg.  The reason for this is simple.
Because of the steep density gradient at the outer edge of the core,
some mass ejection from there is likely, provided the core-bounce
shock is sufficiently strong. This matter has been at high densities,
but because of neutrino phase-space blocking Y$_e$ remains moderately
high until the mass shells pass through the neutrino sphere. At this 
stage, fast deleptonization sets in and produces the low electron 
fractions found by Hillebrandt, Nomoto, \& Wolff. We note that this
mechanism for producing very neutron-rich ejecta is rather independent
of the actual explosion energy but only requires that some core material
is ejected at all.  Woosley \& Baron (1992) argued that neutrino capture 
above the neutrinosphere will tend to raise the value of Y$_e$.  This 
is a very model-dependent process, but deserves more study.    

The other calculations of collapsing degenerate cores have been
summarized by Colgate, Fryer, \& Hand (1997), who conclude that
the results depend rather sensitively on neutrino source terms
and the equation of state.  Colgate et al. conclude that all similar
models, including their own that invoke the Baron, Cooperstein,
\& Kahana equation of state (e.g. Baron et al 1987a,b; Woosely \& Baron 1992)
fail to get a prompt explosion.  The calculation of Colgate et al.
using a Lattimer-Swesty equation of state 
(Lattimer \& Swesty 1991) did yield an 
explosion with 0.08 - 0.2 \Msun\ ejected, very similar to the results
of Hillebrandt, Nomoto, \& Wolff who used an equation of state
similar to that of Lattimer and Swesty.  The issue of 
explosions
in these cores seems to be, at least, open.  Colgate et al. raise
a concern with the overproduction of elements such as $^{88}$Sr in
the neutron-rich ejecta, but this may be suppressed in favor of
the {\it r}-process elements at the very low values of Y$_e$ suggested
here.

The issue for now is whether the collapsing O/Ne/Mg cores of stars
of 8-10 \Msun\ might make reasonable
candidates for the {\it r}-process in a prompt explosion or in the
subsequent $\nu$-driven wind.

\section {The {\it r}-Process}

Considerations of the systematics of the {\it r}-process have
been considerably enhanced by parameter studies of the
conditions necessary to produce the {\it r}-process.  In particular,
the work of Qian \& Woosley (1996) and of Hoffman, Woosley, \& Qian (1997) 
giving analytic and numerical models of the $\nu$-driven wind process
in terms of basic parameters of collapse calculations has
provided important new tools to explore various scenarios, including
the one we propose here.
Hoffman et al. have explored a range of parameters for
the entropy, S, the electron fraction Y$_e$, the dynamic timenscale,
and, for the $\nu$-driven wind process, the mass-loss rate.
Qian \& Woosley have given analytic estimates of the entropy,
mass-loss rate, net neutrino heating, dynamical time scale,
and final electron fraction as a function of neutron star 
mass, radius, neutrino luminosity and neutrino mean energy.
The {\it r}-process parameters
predicted in the prompt explosion of Hillebrandt, Nomoto, \& Wolff (1984)
occupy an extreme in low S and 
low
Y$_e$ in the study of Hoffman et al.  
This regime is close to that in classic studies of the {\it r}-process
(Seeger, Fowler, \& Clayton 1965;  Hillebrandt, Takahashi, \& Kodama 1976).  

The analytic models of Hoffman et al. show an interesting feature that is 
certainly relevant to the scenario we explore here and may be
relevant to the {\it r}-process in general.  They
contain a critical point at the value of Y$_e$ that corresponds
to the ratio $\bar{Z}$/A where $\bar{Z}$ is the mean atomic weight
of the initial seed nucleus and A is the atomic weight of the
target nucleus.  For instance, the denominator of 
equation 19a of Hoffman et al. blows up for Y$_e$ = $\bar{Z}/$A,
formally giving S = 0. This criterion on Y$_e$ is equivalent
to stipulating that the final mass fraction of $\alpha$-particles
is zero and that the initial composition distribution contains
precisely the right fraction of $\alpha$-particles and neutrons
so that they are just used up as the target nucleus is formed.
The critical point in the expressions of Hoffman et al. should
not be taken literally, but it serves as a reminder that
a choice of the right initial conditions helps to make the final
product more easily (Seeger, Fowler \& Clayton 1965).
This may suggest that nature is looking
for an {\it r}-process mechanism for which the initial value of 
Y$_e$ is nearly equal to the ``target" value of $\bar{Z}/$A.

More generally, the entropy required to make platinum is a sensitive 
function of Y$_e$ for low values of Y$_e$.  From equation 19a of Hoffman
et al. we find that for a low Y$_e$ $\sim$ 0.2 and a dynamical 
time scale $t_{dyn,-3}\sim 0.5$ in units of $10^{-3}$s 
for the prompt explosion of Hillebrandt et al. that 
\begin{equation}
S = 200 t_{dyn,-3}^{1/3}(Y_{e,i} - \bar{Z}/A)^{2/3}.
\end{equation}
\noindent
For $\bar{Z}$ = 34 and A = 190, or $\bar{Z}/$A = 0.17,
equation 1 gives S =30 for Y$_{e,i}$ = 0.2 and S = 15,
the value predicted for the prompt explosion of the
model of Hillebrandt et al., for Y$_{e,i}$ = 0.19.  Note
that this model (and equation 19a from which it was derived)
cannot handle the case of  Y$_{e,i}$ = 0.16, which is in
the range quoted by Hillebrandt et al.
These considerations suggest that the sort of collapse
predicted by Hillebrandt et al. is a qualitatively viable 
candidate for the {\it r}-process and should be examined more
rigorously.

Whether in such a model the envelope is ejected in a prompt shock
or the shock stalls,
neutrino emission from the newly-born cooling neutron star will drive
a wind similar to what has been found in case of collapsing iron cores. 
Such a wind, lasting for several seconds, will also blow
off the outer layers of the proto-neutron star even 
if the shock stalls, as it will for more massive stars.  
Because the neutron stars formed in this process are
significantly different in several ways, one expects the
{\it r}-process in the associated $\nu$-driven wind to be quantitatively
different than that previously considered. It is
beyond the scope of this Letter to explore this subject in
any detail and indeed many of the relevant parameters are
not readily available.  We will just take note of some
qualitative effects with the aid of the analytic models
of Qian \& Woosley (1996).

Qian \& Woosley show that the final entropy expected in a $\nu$-driven
wind is not very sensitive to the key parameters. It depends
linearly on the neutron star mass, M, and on the radius of
the neutron star as R$^{-2/3}$, provided mean neutrino luminosity and 
energy remain unchanged. A lower mass neutron star
will have a somewhat larger radius, so these factors will
both decrease somewhat as a result of decreasing the mass of the 
neutron star.  The expected increase in the
radius, of order 30 to 50 percent, is within
the current uncertainties of estimates of neutron star radii
for different equations of state.  The effect will, in any case, 
not be large. The assumption on which this conclusion rests,
namely that less massive neutron stars will radiate neutrinos
with essentially the same mean energy, is less certain.
It may well be that for low-mass neutron stars with more extended
surface layers, the mean neutrino energy is lower than for more massive
ones, again reducing the final entropy somewhat. The most likely
change in the final entropy will be a small reduction, probably 
by less than a factor of 2.   

The mass-loss rate in the wind scales more sensitively with
all the parameters, going as 
$L_{\nu}$$^{5/3}$$\epsilon_{\bar\nu_e}$$^{10/3}$ M$^{-2}$R$^{5/3}$.  
The net neutrino heating also depends rather sensitively on
these parameters, going as
$L_{\nu}$$^{5/3}$$\epsilon_{\bar\nu_e}$$^{10/3}$M$^{-2}$R$^{2/3}$. 
The dynamical time scale goes as
$L_{\nu}$$^{-1}$$\epsilon_{\bar\nu_e}$$^{-2}$MR$^{-1}$. 
The final value of $Y_e$ will be that at the last
achievable NSE value (Qian \& Woosley 1996) and will
also depend on these parameters in a way that must be 
determined.   

The expectation is, to first order, that the neutrino luminosity
will not be substantially different for 
the lower mass neutron stars we consider here. Because
the binding energy to be liberated is less by about
50 percent, the same neutrino luminosity would imply
a shorter Kelvin-Helmholtz cooling time.  The shorter
cooling time coupled with 
the lower entropy and mean neutrino energies would tend to work 
against the {\it r}-process in a $\nu$-driven wind, but we do not know
the difference between the ${\nu_e}$ and ${\bar\nu_e}$ energies, which
determines the final electron fraction.
The steeper density jump at the interface between the
neutron star surface and the outer envelope might
give a larger entropy from the start and thus a more 
effective {\it r}-process in the wind. These effects will tend to balance
each other, and the {\it r}-process in the ${\nu}$-driven wind
may be similar for massive stars and those considered here.

\section {Conclusions }

We have shown that a convergence of arguments
points to the collapsing O/Ne/Mg cores of stars
in the mass range 8 to 10 \Msun\ as a promising
site of the {\it r}-process.  This mass range should
produce neutron stars by core collapse and has
been generally understudied to this point.
The progenitor evolution is quite complex with multiple
shell flashes and active electron capture.  The collapse
is expected to be quantitatively if not qualitatively
different than for iron cores.

If these collapses produce an 
explosion of
the kind found by Hillebrandt, Nomoto, \& Wolff (1984),
then this explosion with a rather low value of the
initial electron fraction could produce a substantial
{\it r}-process with the entropies predicted in the model.
If the neutrino luminosity and mean neutrino
energy are not too different from those obtained
for collapsing iron cores, then the differences in the
parameters determining the subsequent $\nu$-driven wind
{\it r}-process may not be radically
different for the collapsing O/Ne/Mg cores.  This may
mean that all $\nu$-driven wind processes can
contribute substantially to the first and second {\it r}-process
peak, but that a special characteristic of the low
mass cores, e.g. a prompt explosion, may give the bulk of
the third peak. Alternatively, these low-mass supernovae 
explosions could be responsible for the production of all of
the {\it r}-process elements above Z=55.

A possible objection to this
scenario arises from the observed propensity of 
neutron stars, at least those in binary systems where their
mass can be measured, to cluster closely around a 
gravitational mass of 1.4 \Msun.  Their proper mass must be somewhat larger.
This requires some special circumstances if this homogeneity is to be
taken literally.  We note that the sample of neutron stars with
measured masses is small and that neutron stars that arise from
collapsing O/Ne/Mg cores may be a relatively small fraction
of all neutron stars. Unless one believes that all stars from
8 to 10 \Msun\ lose their envelopes and that their cores
never collapse, these stars must produce neutron stars.  We acknowledge
that no such neutron star has ever been directly recognized,
but feel that these stars are such a fruitful and overlooked
site for the {\it r}-process that it is worth giving them closer
scrutiny. 

The observational
indication that neutron stars originating from this mass range are
rare may support our case.
To estimate the total contribution of a given class of models to
the Galactic {\it r}-process, one must have some idea of the event
rate.  For a Salpeter mass function, stars from 8 to 10 \Msun\
represent a fairly large fraction of all stars with mass 
in excess of 8 \Msun, of order 0.3.  For a somewhat steeper
Miller-Scalo mass function, this fraction would be somewhat higher.
Estimates of the {\it r}-process give an ejecta mass of order
$10^{-4}$ \Msun\ per event for all supernovae (Hoffman, Woosley, \& Qian
1997).  
If collapsing O/Ne/Mg cores were to eject 0.02 \Msun\ of {\it r}-process
elements
per event (neglecting for the moment a subsequent $\nu$-driven wind)
and they represent over 10 percent of all supernovae,
then there is a strong possibility of overproducing the
{\it r}-process.  This is a liability in some way, but also gives
ample room for inefficiency. 

One obvious source of inefficiency is related to the question
of envelope loss. A radiative wind or ``superwind" operating 
at $\sim$ $10^{-4}$ \Msun\ yr$^{-1}$ could remove the whole
envelope and teminate the growth of the degenerate core before
collapse ensues  (Garc\'ia-Berro, Ritossa, \& Iben 1997:
and references therein).  The systematics of this mass loss from
stars of 8 - 10 \Msun\ are unknown and worth futher study.  Another
source of inefficiency is in the collapse calculations themselves. 
The calculations of Hillebrandt et al.
under-resolved the outer layers of the core and so the mass ejected
with very low Y$_e$ was probably too high. Moreover, they did not
include the effects of thermal neutrino losses, which would weaken the 
prompt shock, thereby also reducing the mass ejected. Finally, an
equation of state was used which also favored prompt mass ejection. 
In conclusion, the amount of low Y$_e$ material ejected from the
collapsed O/Ne/Mg core could in reality be considerably less than was 
found by Hillebrandt et al. We note, however, that if prompt explosions
were to occur in all supernovae and to produce the {\it r}-process,
there would be a definite problem with overproduction (Hillebrandt 1978;
Mathews \& Ward 1985).
This is another argument for producing the {\it r}-process in only
a special category of supernovae, and some collapsing O/Ne/Mg cores might
resemble such a class. 

The primary conclusion of this study is that the
evolution and collapse of cores of stars of 8 to 10 \Msun\ 
and the associated {\it r}-process deserves much greater quantitative study.

\section*{ACKNOWLEDGMENTS} 

The authors thank the Institute for Theoretical Physics
at the  University of California at Santa Barbara, 
where this work was catalyzed.  They are grateful for insightful conversations
on these topics with Rob Hoffman, Ken Nomoto, and Adam Burrows.   
The ITP is supported by NSF under Grant No. PHY94-07194.
Other support for this research was provided by NSF Grants AST-9528110,  
AST-9314936 and AST-9618332, NASA Grant NAG 5-2888, grants GO-05421, GO-05856,
and GO-06748 from the Space Science Telescope Institute, which is
operated by the Association of Universities for Research in Astronomy,
Inc., under NASA contract NAS5-26555, by the SFB 375 ``Astro-Teilchenphysik''
of
the Deutsche Forschungsgemeinschaft and by a grant from the 
Texas Advanced Research Program

\end{document}